\documentclass[prl,twocolumn,superscriptaddress]{revtex4}

\usepackage{graphicx}
\usepackage[normalem]{ulem} %need for strikethrough
\usepackage{color} %needed for coloring of \todo etc.

\begin{document}

\title{Anisotropic rare-earth spin ensemble strongly coupled to a superconducting resonator}

\author{S.~Probst}
\affiliation{Physikalisches Institut, Karlsruhe Institute of Technology, D-76128
Karlsruhe, Germany}
%\affiliation{DFG-Center for Functional Nanostructures (CFN), D-76128 Karlsruhe, Germany}

\author{H.~Rotzinger}
\affiliation{Physikalisches Institut, Karlsruhe Institute of Technology, D-76128
Karlsruhe, Germany}

\author{S.~W\"{u}nsch}
\affiliation{Institut f\"{u}r Mikro- und Nanoelektronische
Systeme, Karlsruhe Institute of Technology, D-76189 Karlsruhe, Germany}

\author{P.~Jung}
\affiliation{Physikalisches Institut, Karlsruhe Institute of Technology, D-76128
Karlsruhe, Germany}

\author{M.~Jerger}
\affiliation{Physikalisches Institut, Karlsruhe Institute of Technology, D-76128
Karlsruhe, Germany}

\author{M.~Siegel}
\affiliation{Institut f\"{u}r Mikro- und Nanoelektronische
Systeme, Karlsruhe Institute of Technology, D-76189 Karlsruhe, Germany}

\author{A.~V.~Ustinov}
\affiliation{Physikalisches Institut, Karlsruhe Institute of Technology, D-76128
Karlsruhe, Germany}

\author{P.~A.~Bushev}
\affiliation{Physikalisches Institut, Karlsruhe Institute of Technology, D-76128
Karlsruhe, Germany}

\date{\today}

\begin{abstract}
Interfacing photonic and solid-state qubits within a hybrid quantum architecture offers a promising route towards large scale distributed quantum computing. Ideal candidates for coherent qubit interconversion are optically active spins magnetically coupled to a superconducting resonator. We report on a cavity QED experiment with \emph{magnetically anisotropic} Er$^{3+}$:Y$_2$SiO$_5$ crystals and demonstrate strong coupling of rare-earth spins to a lumped element resonator. In addition, the electron spin resonance and relaxation dynamics of the erbium spins are detected via direct microwave absorption, without aid of a cavity.
\end{abstract}
\maketitle

%\pacs{42.50.Fx, 76.30.Kg, 03.67.Hk, 03.67.Lx, 76.30.-v}
%\keywords{Cooperative phenomena in quantum optical systems, Superconducting qubits, Quantum communication, Quantum computation architectures and implementations, EPR in condensed matter}

Quantum communication networks are considered to distribute entangled states over a large scale computing architecture~\cite{Cirac1997,Kimble2008}. The core elements of future quantum networks, i.e. quantum repeaters~\cite{Gisin2007} as well as network nodes, will be realized by using qubits and quantum memories of diverse physical nature~\cite{Tian2004, Rabl2010}. Today, elementary quantum networks linking two remote single atoms have been demonstrated~\cite{Rempe2012, Weinfurter2012}. Solid state systems, such as superconducting~(SC) quantum circuits~\cite{Clarke2008}, nanomechanical devices~\cite{Cleland2010} and spin doped solids~\cite{Wu2010} potentially offer larger scalability and faster operation time compared to systems based on the single atom approach. However, such solid state devices operate at microwave and RF frequencies, which are less suitable for long-range quantum communication than optical channels due to losses in cables and the high noise temperature of antennas. Therefore, to establish a fiber-optical link between them, one has to use a quantum media converter, i.e. a device which coherently interfaces matter and photonic qubits~\cite{Lukin2010, Tittel2011, Gisin2011, Yamamoto2012}.

One of the promising ways towards implementation of such a converter relies on using optically active spin ensembles in a hybrid quantum architecture~\cite{Molmer2008,Verdu2009,Bushev2011}. At the present time, the research activity is primarily focused on various cavity QED experiments with spin ensembles of NV-centers in diamond~\cite{Bertet2010,Shuster2010,Amsuss2011,Zhu2011,Bertet2011}. Recently, collective strong coupling has been demonstrated in conventional electron spin resonance~(ESR) experiments with organic molecules in a 3D cavity~\cite{Chiorescu2010,Morton3D}, and four-wave mixing from Fe$^{3+}$ ions in a 3D sapphire loaded cavity has been shown~\cite{Tobar2012}. Surprisingly, less progress has been achieved with paramagnetic laser materials such as ruby~\cite{Shuster2010} and rare-earth ions doped crystals~\cite{Bushev2011,Wilson2012}. The way to the strong coupling regime in these initial experiments was impeded by the large inhomogeneous broadening of spin ensembles~\cite{Molmer2011}.

Unlike electronic spins of 3$d$ transition series metal ions, Kramers rare-earth~(RE) ions reveal a rather strong magnetic anisotropy due to the distortion of their 4$f$ electronic orbitals by a crystal field, see Chapter 5 of Ref.~\cite{AbragamESR}. Such a distortion induces a strong dependence of the g-factors and the Rabi-frequency on the orientation of the RE-ion doped crystal with respect to the polarizations of the permanent~(DC) and the oscillating~(AC) magnetic fields~\cite{Bertaina2009}. The magnetic moments of Er$^{3+}$ ions in certain crystals with axial symmetry can vary from zero to 7.5$\mu_B$, where $\mu_B$ is the Bohr magneton~\cite{Guillot2006, Sun2008}. Using such \emph{magnetically anisotropic} materials in hybrid quantum systems leads to new challenges. In this letter, we present cavity QED experiments on anisotropic Er$^{3+}$:Y$_2$SiO$_5$~(Er:YSO) crystals and demonstrate strong collective coupling of the RE-spin ensemble to a superconducting lumped element~(LE) resonator.

\begin{center}
\begin{figure*}[htb]
\includegraphics[width=1.5\columnwidth]{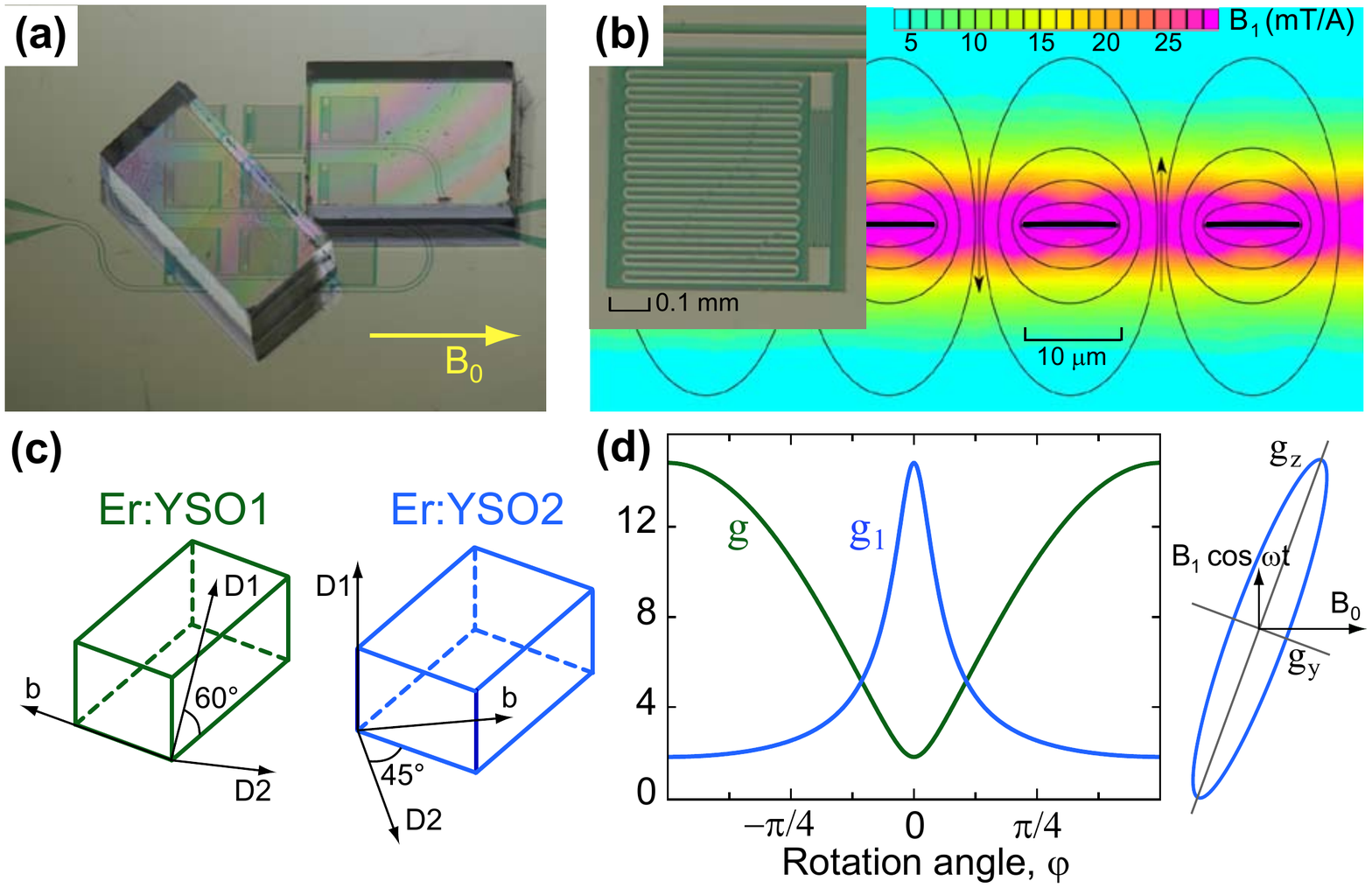}
\caption{(Color online) \textbf{(a)} Picture of the experiment: two Er:YSO crystals are placed on the SC-chip with 9 LE resonators. The DC magnetic field is applied along the chip's surface. \textbf{(b)} The simulated AC magnetic field in the plane perpendicular to the DC field in the vicinity of the LE inductor (meander structure). (Inset) Picture of a LE resonator. \textbf{(c)} Orientation of the crystal's axis. \textbf{(d)} Dependence of $\textrm{g}$ and $\textrm{g}_1$ on the rotation angle of the crystal around the x-axis of the g-tensor. The maximal coupling is reached when the AC field is aligned along the largest component of the g-tensor.}\label{Setup}
\end{figure*}
\end{center}

The experimental setup is shown in Fig.~\ref{Setup}(a). Two Er:YSO crystals are glued on a superconducting niobium chip containing 9 lumped element resonators coupled to a 50~$\Omega$ transmission line~\cite{Wuensch2011}. The resonance frequencies of the circuit occupy the microwave band between 4.5 and 5.2~GHz and exhibit loaded and intrinsic quality factors of $Q_c\simeq$~10$^3$ and $Q_i\simeq$~10$^4$, respectively. The quality factors of the LE resonators do not degrade significantly by applying an in-plane DC magnetic field up to 340~mT. Figure~\ref{Setup}(b) presents a simulation of the AC magnetic field and the picture of a single LE resonator. Due to the narrow width of 10~$\mu$m of the meandered inductance, the AC microwave field $B_1 \cos \omega t$ is mainly concentrated in direct vicinity of the chip. The mode volume of such a resonator is $V_m \approx$~10$^{-6}$~cm$^{3}$ comprising $N_s\sim$~10$^{12}$ spins for magnetic coupling.

We study two Y$_2$SiO$_5$ crystals~(noted below as Er:YSO1 and Er:YSO2) doped with 200~ppm Er$^{3+}$ ions (Scientific Materials, Inc.). The orientation of their optical extinction axes $b$, $D_1$ and $D_2$ is sketched in Fig.~\ref{Setup}(c)~\cite{Sun2008}. The alignment of the crystals in the DC magnetic field $\vec{B}_0$ is specified by the angles $\theta$ and $\phi$. The comparison of measured ESR spectra to the ``Easyspin'' simulation~\cite{EasySpin} yields the following pair of angles $\theta_1=$~44$^{\circ}$, $\phi_1=$~111$^{\circ}$ and $\theta_2=$~48$^{\circ}$, $\phi_2=$~85$^{\circ}$ for the Er:YSO1 and Er:YSO2 crystals, respectively.

Figure~\ref{Setup}(d) qualitatively outlines the main idea of the experiment: the Er:YSO crystal is known to have the strongest magnetic anisotropy due to its low axial symmetry $C_{2h}$, and the principal values of the g-tensor for the crystallographic site 1 are $\textrm{g}_x\approx$~0, $\textrm{g}_y=$~1.5 and $\textrm{g}_z=$~14.8, see~\cite{Guillot2006}. The angular dependence of the DC and AC \textrm{g}-factors ($\textrm{g}$ and $\textrm{g}_1$) of the crystal rotated around $x$-axis of the \textrm{g}-tensor are given by (s.f. Chapter 3 of Ref.~\cite{AbragamESR})
\begin{eqnarray}\label{gfactors}
\textrm{g}^2(\varphi) &=& \textrm{g}_y^2\cos^2\varphi+\textrm{g}_z^2\sin^2\varphi, \\
\textrm{g}_1(\varphi) &=& \textrm{g}_y\textrm{g}_z/\textrm{g}(\varphi).
\end{eqnarray}

The coupling strength of a single spin to a microwave photon with an AC field of $\vec{B}_1 \cos \omega t$ is determined by $v_1 = \mu_B \textrm{g}_1(\varphi) |\vec{B_1}|/2\hbar$. Therefore, to attain the largest coupling strength, the AC field has to be polarized along the $\textrm{g}_z$ direction and the DC field should be aligned perpendicular to the AC field. In the following we show, that in order to reach the strong coupling regime one has to sacrifice the high spin tuning rate of Er:YSO.

\begin{center}
\begin{figure*}[htb]
\includegraphics[width=1.5\columnwidth]{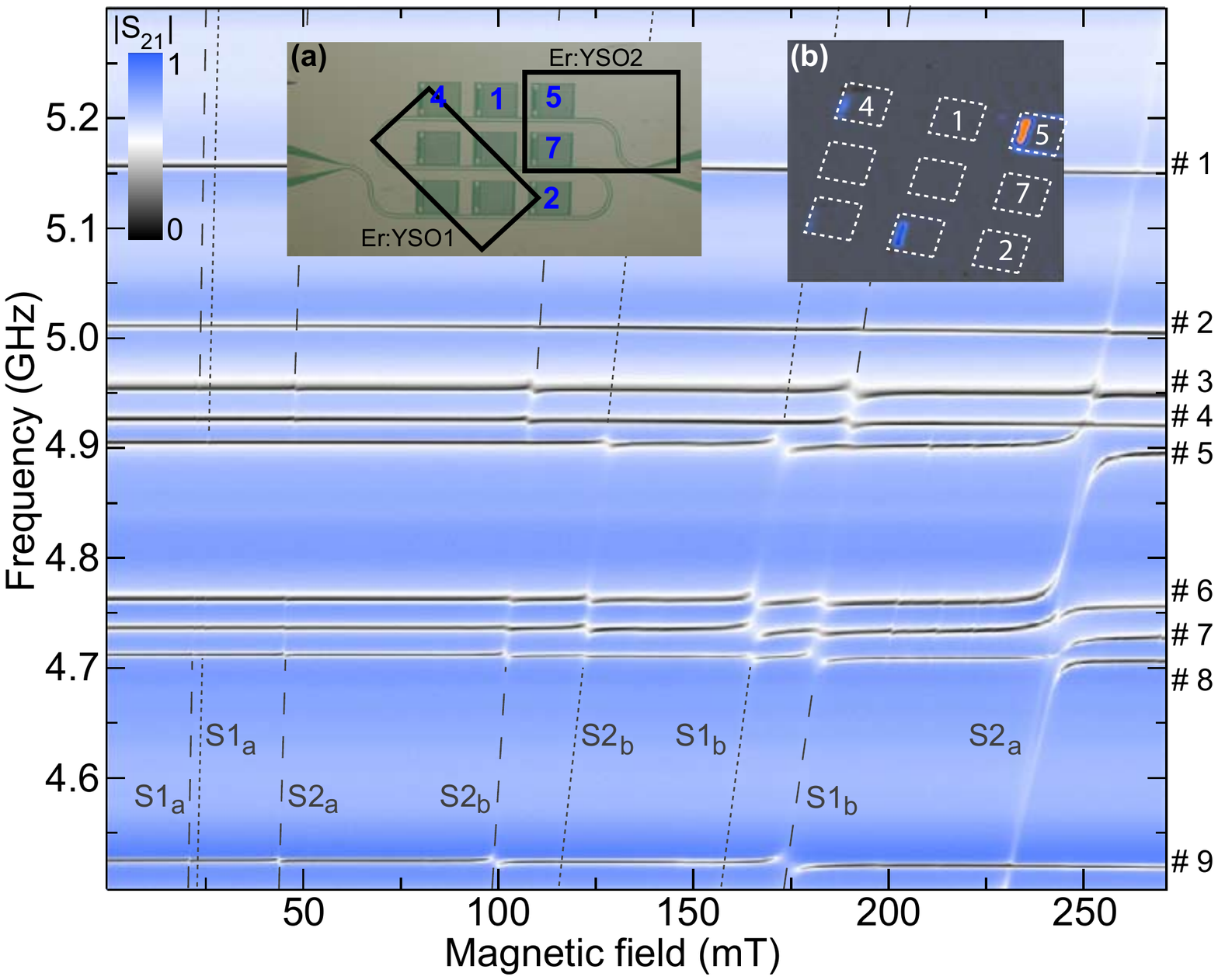}
\caption{(Color online) ESR spectrum of the Er:YSO crystals coupled to the LE chip. The dashed lines correspond to 4 electronic spin transitions of the Er:YSO1 crystal and the dotted lines to transitions of the Er:YSO2 crystal. The resonator labels are given on the right. (Inset a) Positions of the LE resonators beneath the crystals. (Inset b) LSM reflectivity map measured at the excitation frequency of resonator \#5.}\label{LargeSpec}
\end{figure*}
\end{center}

The experimental system is placed inside a BlueFors LD-250 dilution fridge and the on-chip ESR spectroscopy is performed in a temperature range of 20-500~mK. Figure~\ref{LargeSpec} presents the transmission spectrum of the circuit $|S_{21}(\omega)|$ as a function of the applied magnetic field. The excitation power at the input of the circuit is approximately 50~aW corresponding to a few microwave photon excitation level inside each resonator. Every horizontal line in the ESR spectrum corresponds to one of the 9 LE resonators on the chip.  Below 200~mT, the resonance lines are interrupted by a regular pattern of dispersive cavity shifts due to the coupling to the magnetic dipole transitions of Er$^{3+}$ between the electronic states $m_s=\pm$1/2.

In general, the ESR spectrum of Er:YSO consists of two pairs of strong lines associated with the occupation of Er$^{3+}$ ions of two distinct crystallographic sites and each in two magnetically inequivalent positions~\cite{Kurkin1980,Guillot2006,Sun2008}. The dashed and dotted lines in Fig.~\ref{LargeSpec} correspond to the spin tuning lines of the samples Er:YSO1 and Er:YSO2, respectively. The term $S1_a$, for instance denotes the transition of the erbium ions in site 1 and magnetic inequivalent position~(magnetic class) $a$. Correspondingly, the term $S2_b$ belongs to site 2 and magnetic class $b$. The effect of the magnetic anisotropy is clearly seen in the spectrum in Fig.~\ref{LargeSpec}: High field transitions with smaller $\textrm{g}$-factor have larger coupling strength.

Inset~(a) in Fig.~\ref{LargeSpec} shows the positions of the resonators beneath the crystals which were partially identified by measuring the local photoresponse of the circuit in a laser-scanning-microscope~(LSM) setup~\cite{Zhuravel2006}. A typical LSM reflectivity map taken at the excitation frequency of resonator \#5 is presented in inset~(b). These reflectivity maps have been taken during a separate run at 4~K, and only an incomplete identification of the resonators was possible.

In the field region between 210 and 265~mT, a faint narrow absorption line traverses the spectrum, which corresponds to the $S2_a$ transition of the Er:YSO2 crystal. The interception of this line with resonances \#6-8 results in a complex hybridization pattern, due to the mutual inductive coupling between those resonators.

\begin{center}
\begin{figure*}[htb]
\includegraphics[width=1.5\columnwidth]{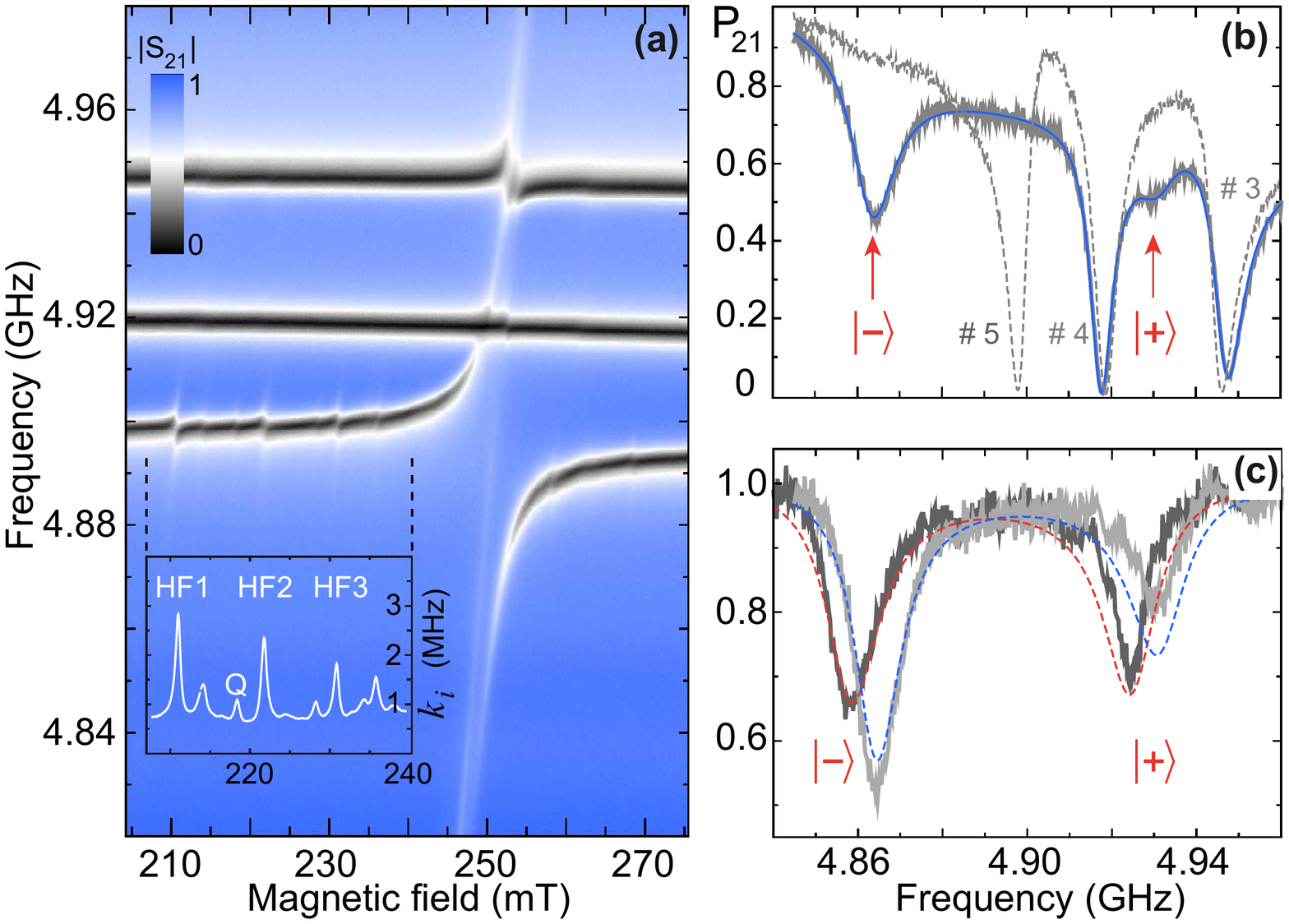}
\caption{(Color online) \textbf{(a)}~Transmission spectrum of resonator \#5 strongly coupled to the electronic spin $S2_a$ transition. (Inset)~The internal damping rate of the resonator $k_i/2\pi$ clearly resolves the hyperfine spectrum of $^{167}$Er$^{3+}$. The "HF's" denote magnetic hyperfine transitions and "Q" marks a quadrupole transition. \textbf{(b)}~The gray dashed line is the transmitted power $P_{21}=|S_{21}(\omega)|^2$ measured away from the avoided level crossing at 204.3~mT. The dark gray line displays the power spectrum at the anticrossing at 251.3~mT, which is fit using Eq.~\ref{BaselineCorrection}~(blue line). \textbf{(c)}~The corrected transmitted power spectra taken at 250.6~mT (dark gray line) and at 251.3~mT (gray line) clearly show a normal mode splitting. The dashed lines show the fit to this splitting.}\label{AvoidedLevelCrossing}
\end{figure*}
\end{center}

Figure~\ref{AvoidedLevelCrossing} presents the detailed study of the avoided level crossing between the erbium transition $S2_a$ and resonator \#5 at 250.6~mT. The raw transmission spectrum $|S_{21}(\omega)|^2$ of the circuit measured at the avoided level crossing at 253.1~mT is shown in Fig.~\ref{AvoidedLevelCrossing}(b). For comparison, we show the transmission spectrum taken away from the anticrossing at 204.3~mT. A typical experimental power spectrum $|S_{21}(\omega)|^2$ measured at the center of the avoided level crossing consists of the normal mode splitting itself, 2 additional uncoupled resonances and the absorption dip. The spectrum $|S_{21}(\omega)|^2$ can be described by the following expression
\begin{equation}\label{BaselineCorrection}
|S_{21}(\omega)|^2 = B(\omega)\left[1+ \sum_{i=1}^5~\frac{a_{1i}+a_{2i}(\omega-\omega_i)}{(\omega-\omega_i)^2+\gamma_i^2/4}\right],
\end{equation}
where $B(\omega)$ is a second order polynomial, accounting for the baseline, $\omega_i$ are the frequencies of the resonances and $\gamma_i$ are their linewidths. The coefficients $a_{1i}$ and $a_{2i}$ describe the absorptive and the dispersive parts of the resonances. The fit of the experimental data recorded at 251.3~mT to Eq.~\ref{BaselineCorrection} is presented in Fig.~\ref{AvoidedLevelCrossing}(b).

The dips marked with $|+\rangle$ and $|-\rangle$ correspond to the normal mode splitting. The \#3 and \#4 denote the positions of the resonators which are not coupled to Er:YSO2. In order to extract the spectrum of the normal mode splitting we normalize the measured curve $|S_{21}(\omega)|^2$ to the baseline $B(\omega)$ and subtract the non-interacting resonators \#3 and \#4 as well as the transmission line absorption dip. Figure~\ref{AvoidedLevelCrossing}(c) shows the corrected spectra measured at field values of 251.3~mT~(gray line) and at 250.6~mT~(dark gray line). Two well-separated dips are confirming strong coupling of the spin ensemble $S_{2a}$ to the cavity. To describe the normal mode splitting we use the model of coupled quantum harmonic oscillators~\cite{Verdu2009,Henschel2010,Bertet2010}. The corrected spectra are fit to the avoided level crossing curves according to the equations given in Ref.~\cite{Shuster2010} and presented in Fig.~\ref{AvoidedLevelCrossing}(c) by dashed lines. The external $k_c/2\pi=$~4.7~MHz and internal $k_i/2\pi=$~0.7~MHz damping rates of resonator \#5 were determined away from the avoided level crossing at a field of 204.3~mT and set constant for the fit. We obtain a splitting size of $2v/2\pi=$~68$\pm$1~MHz and a spin linewidth (FWHM) of $\Gamma_2^{\star}/\pi=$~24$\pm$0.5~MHz, which gives the cooperativity parameter $C~=~2v^2/\kappa_c\Gamma_2^{\star}\approx$~36.

The inset in Fig.~\ref{AvoidedLevelCrossing}(a) shows the influence of the $^{167}$Er$^{3+}$ hyperfine transitions on the damping rate of resonator \#5. Particularly at frequencies below 6~GHz the hyperfine spectrum is rather complex and consists of about 20 lines in the field range of our interest (205-280~mT). We could not identify all magnetic hyperfine transitions of Er:YSO with sufficient confidence. Above 6~GHz the hyperfine spectrum appears as a regular pattern of lines and the classification of the spin transitions is easier~\cite{Guillot2006, Bushev2011}. Nonetheless, the transitions between states with equal nuclear spin projection $\Delta m_I=$~0~(HF in the inset) possess larger coupling strengths than quadrupole transitions with $\Delta m_I=\pm$1 (Q in the inset). The fit of the damping rate of the cavity $k_i/2\pi$, weakly coupled to HF1 transition, yields $v_{hf1}/2\pi=$~4~MHz coupling and a linewidth of $\Gamma_{hf1}^{\star}/2\pi=$~7.6~MHz~(see Ref.~\cite{Bushev2011} for the fitting procedure). For the quadrupole transition Q the coupling strength of spins and their linewidth are $v_{q}/2\pi=$~1.8~MHz and $\Gamma_{q}^{\star}/2\pi=$~7.1~MHz respectively.
\begin{table}[htb]
\begin{tabular}{|c||c|c|c||c|c|c|}
  \hline
  Transition & $\textrm{g}$-factor & $v/2\pi~$ & $\Gamma_2^{\star}/2\pi$ & $\textrm{g}$-factor & $v/2\pi~$ & $\Gamma_2^{\star}/2\pi$ \\ \hline
    $S_{1a}$ & 15.2 & 4.1 & 20 & 13.4 & 4.4 & 23 \\ \hline
    $S_{2b}$ &  7.3 & 8 & 28 & 2.7 & 13 & 34 \\ \hline
    $S_{1b}$ & 3.3 & 13 & 24 & 1.6 & 22 & 32 \\ \hline
    $S_{2a}$ & 1.8 & 21 & 26 & 1.4 & 34 & 12 \\ \hline
  \end{tabular}
  \caption{\textrm{g}-factors, coupling strengths $v$ and linewidths $\Gamma_2^{\star}$ in MHz of 4 spin transitions of the Er:YSO1 coupled to resonator \#9~(left), and Er:YSO2 coupled to resonator \#5~(right).}\label{TableYSO2}
\end{table}

Low-field electronic spin transitions of the Er:YSO1 and the Er:YSO2 crystals do not reveal the strong coupling; their coupling strengths and linewidths are found from dispersive shift of the cavity. The results are summarized in the Table~\ref{TableYSO2}. The coupling strength is inversely proportional to the g-factor. The spin linewidth for the $S_{2a}$ transition for Er:YSO2 drops to 12~MHz after an initial increase. The study of Er:YSO2 crystal on the X-band ESR spectrometer Bruker Elexsys 580 at a temperature of 6.3~K suggests a minimal linewidth $\Gamma^{\ast}_2/2\pi\simeq$~14~MHz for low field transitions and $\Gamma^{\ast}_2/2\pi\simeq$~25~MHz for the high field ones. The measured behavior of the spin linewidth in the on-chip ESR experiment can be explained by the spatial inhomogeneity of the magnetic field in the vicinity of the SC chip due to the Meissner effect. The angular variations of the magnetic field broadens mainly those transitions which are close to the sharp vertex of the g-factor ellipse, see also Fig.~\ref{Setup}(d).

The measured narrow linewidth of the low-field erbium transition $S2_a$ is consistent with the measurements of the on-chip microwave absorption. In the absence of any resonator, the microwave signal is extinguished just due to the proximity of the spins to the transmission line. Such an effect has recently been demonstrated with a highly doped ruby crystal~\cite{Shuster2010}. In contrast, the spin concentration in our experiment is $ n_{Er} \simeq $~7$\times$10$^{17}$~cm$^{-3}$, which is at least by a factor of 10 lower. A clear signal results from microwave absorption by $N_s\sim$~10$^{13}$ spins with large AC g-factor of nearly 15.

%However, we observe a clear signal due to the large magnetic moment of Er$^{3+}$ of nearly 7$\mu_B$.
%

\begin{center}
\begin{figure*}[htb]
\includegraphics[width=1.5\columnwidth]{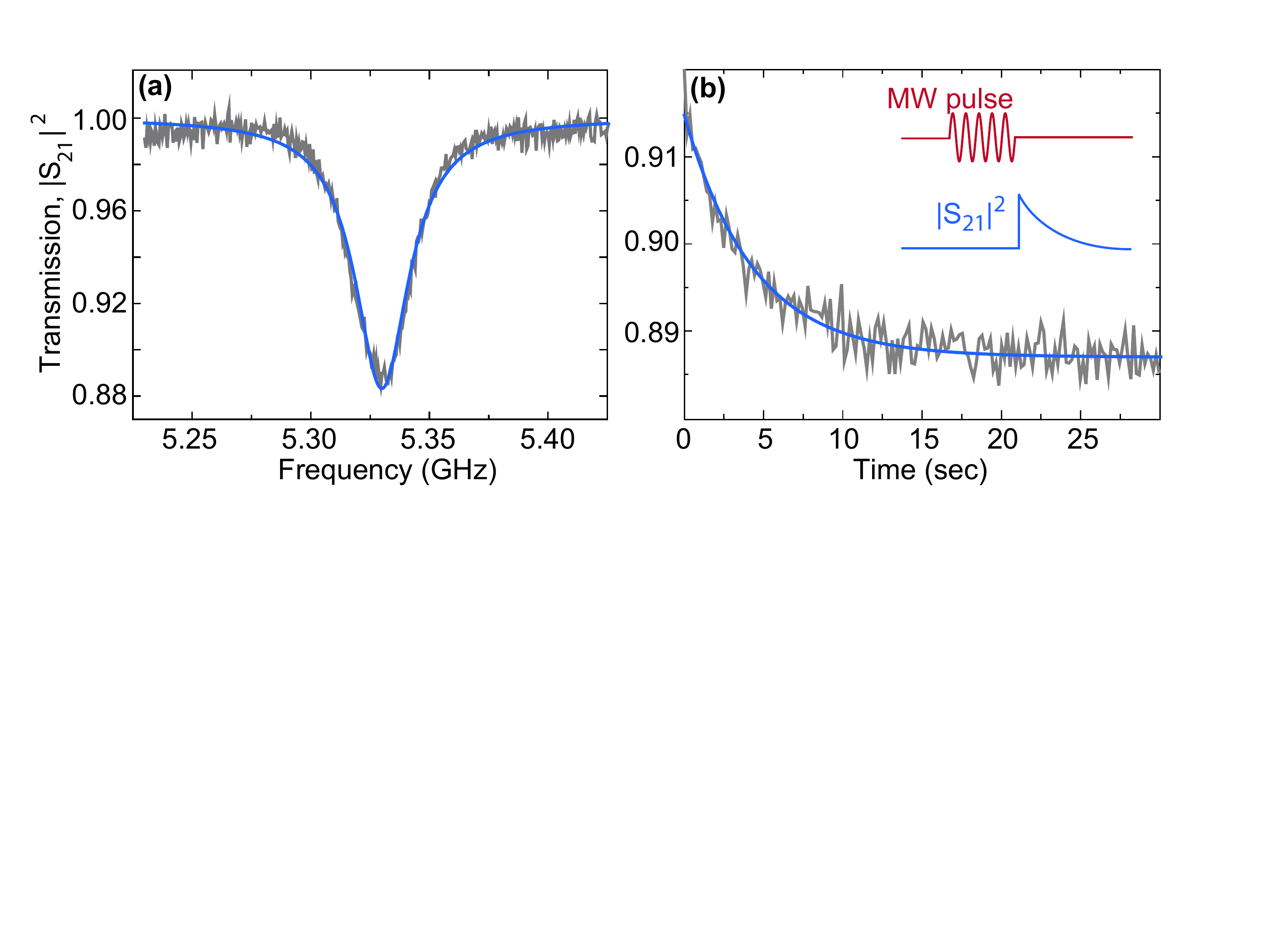}
\caption{(Color online) \textbf{(a)} The gray curve presents the absorption profile of the erbium spins coupled to the transmission line at 273.2~mT. The solid line shows a Lorentzian fit to the data. \textbf{(b)} Reappearance of the absorption signal~(gray curve) at 5.331~GHz after irradiation of the chip with an intense microwave pulse. The solid line is a fit of the data to an exponential decay.}\label{Absorption}
\end{figure*}
\end{center}

We studied the absorption line at a magnetic field of 273.2~mT, which corresponds to an ESR frequency of 5.331~GHz~(440~MHz away from resonator \#5). In order to distill the absorption profile from the transmitted signal $|S_{21}|^2$, we normalized the latter to the far detuned~(140~MHz) transmission taken at 266~mT. The resulting absorption spectrum recorded at an excitation power of 0.1~fW is shown in Fig.~\ref{Absorption}(a). The experimental data (gray dots) show 12\% absorption of the transmitted microwave signal and fit well to a Lorentzian (solid line) with an ESR inhomogeneous linewidth $\Gamma_2^{\star}/2\pi=$~13.8$\pm$0.4~MHz. The linewidth is slightly increased compared to the avoided level crossing due to the larger geometrical size of the transmission line, which makes it more sensitive to the magnetic field inhomogeneity. Therefore, a linewidth narrowing due to the cavity protection effect can be excluded here~\cite{Diniz2011}.

The relaxation time of the spin ensemble at mK temperatures has been recently measured for NV-centers by using the dispersive cavity shift~\cite{Amsuss2011}. In our experiment, the large absorption signal allows for observing the spin relaxation dynamics directly. For that purpose the vector network analyzer was tuned to the absorption dip at $\omega_{a}/2\pi=$~5.331~GHz. The probing power was set to 60~fW and the chip was irradiated for 1~sec with an intense 0.1~nW pulse at a central frequency of 5.330~MHz. The reappearance of the absorption signal is presented in Fig.~\ref{Absorption}(b). The magnitude of the absorption dip is proportional to the average population difference of the spin ensemble, and therefore, to its total magnetization $M_z$
\begin{equation}
|S_{21}|^2(0)-|S_{21}|^2(t)\propto N_1-N_2 \propto M_z(t).
\end{equation}

After the pulse, the population restores its equilibrium level $N_2/N_1= \exp(-\hbar \omega_{a}/k_B T)\sim$~10$^{-5}$, where $T=$~20~mK is the temperature of the experiment. The solid line in Fig.~\ref{Absorption}(b) shows the fit of the measured spin relaxation signal to the exponential decay with $T_1=$~4.3$\pm$0.2~sec.

To conclude, we have presented cavity QED experiments with magnetically anisotropic Er$^{3+}$:Y$_2$SiO$_5$ crystals. The coupling strength and linewidth strongly depend on the g-factors of the erbium spin transitions. The narrow inhomogeneous linewidth of 12~MHz for the high field transition with g~=~1.4 allowed us to attain strong coupling between the SC resonator and the RE-spin ensemble. The ESR as well as the spin relaxation dynamics were detected by direct measurement of the microwave absorption of the spins coupled exclusively to the transmission line. The presented experiment demonstrates the promising potential of rare-earth ion doped crystals for application in hybrid quantum architectures.

We thank Y.~Kubo for the critical reading of the manuscript, S.~Skacel and F.~Song for the technical assistance, I.~Protopopov, G.~Grabovskij and H.~Maier-Flaig for the simulation software. S.~P. acknowledges financial support by the LGF of Baden-W\"{u}rttemberg. This work was supported by the the BMBF Programm “Quantum Communications” through the project QUIMP.

\bibliography{er_lekid}

\end{document}